\documentclass[epj]{webofc}
\usepackage[varg]{txfonts}   
\woctitle{MESON2014 - the 13$^\textrm{th}$ International Workshop on Meson Production, Properties and Interaction}
\begin{document}
\selectlanguage{english}
\title{Charmonium production in $\bar p$-induced reactions on nuclei}

\author{Alexei Larionov\inst{1,2}\fnsep\thanks{\email{larionov@fias.uni-frankfurt.de}} \and
        Marcus Bleicher\inst{1,3} \and
        Albrecht Gillitzer\inst{4} \and
        Mark Strikman\inst{5}
}

\institute{Frankfurt Institute for Advanced Studies (FIAS), 
           D-60438 Frankfurt am Main, Germany
\and
           National Research Centre "Kurchatov Institute", 
           123182 Moscow, Russia
\and
           Institut f\"ur Theoretische Physik, J.W. Goethe-Universit\"at,
           D-60438 Frankfurt am Main, Germany
\and
           Institut f\"ur Kernphysik, Forschungszentrum J\"ulich, D-52425 J\"ulich, Germany
\and
           Pennsylvania State University, University Park, PA 16802, USA
          }

\abstract{%
The production of charmonia in the antiproton-nucleus reactions at $p_{\rm lab}=3-10$ GeV/c is studied within 
the Glauber model and the generalized eikonal approximation. 
The main reaction channel is charmonium formation in an antiproton-proton collision.
The target mass dependence of the charmonium transparency ratio allows to determine the charmonium-nucleon 
cross section. 
The polarization effects in the production of $\chi_{c2}$ states are evaluated.
}
\maketitle
\section{Introduction}
\label{intro}

Investigation of charmonium-nucleon interactions is important for several reasons:
(i) Interpretation of $J/\psi$ suppression in relativistic heavy-ion collisions and separation 
of the charmonium dissociation mechanism in a quark-gluon plasma from 
cold nuclear matter effects.
(ii) Constraining the QCD-inspired models of charmonia.   
(iii) Growing understanding of nonperturbative vs perturbative QCD aspects, such as factorization 
theorem, color dipole cross section, color transparency phenomenon.

Charmonium formation $\bar p p \to R$ ($R=J/\psi, \psi^\prime, \chi_c, \ldots$) on a 
nuclear target proton  has been proposed long ago \cite{Brodsky:1988xz,Farrar:1989vr} to address 
the genuine charmonium-nucleon cross section, since the charmonium formation length is short
in this case. In this talk we present some results of our detailed studies
\cite{Larionov:2013axa,Larionov:2013nga} of $J/\psi, \psi^\prime$ and $\chi_c$ production 
in $\bar p A$ interactions at threshold.

\section{Glauber model calculation of $J/\psi$ and $\psi^\prime$ production}
\label{glauber}

Charmonium production cross section in antiproton-nucleus interaction is calculated as
\begin{equation}
   \sigma_{\bar p A \to R (A-1)^*}= v_{\bar p}^{-1} \int d^2b
   \int\limits_{-\infty}^\infty\,dz\, 
   {\rm e}^{-\sigma_{\bar p N}^{\rm inel}(p_{\rm lab})\int\limits_{-\infty}^z\,dz'\rho(z',b)}
   \Gamma_{\bar p \to R}(z,b)\,
   {\rm e}^{-\int\limits_z^\infty\,dz'\rho(z',b) \sigma_{RN}^{\rm eff}(p_R,z'-z)}~,
                                          \label{sigma_pbarA2R}
\end{equation} 
where $v_{\bar p}=p_{\rm lab}/E_{\bar p}$ is the antiproton velocity.
$\Gamma_{\bar p \to R}$ is the antiproton width with respect to the 
charmonium production:
\begin{equation}
   \Gamma_{\bar p \to R}(z,b)=\int\,\frac{2d^3p}{(2\pi)^3}\, v_{\bar p p}(s)\, 
    \sigma_{\bar p p \to R}(s)\, f_p(z,b,\mathbf{p})~,        \label{Gamma_pbar2R}
\end{equation}
where $f_p(z,b,\mathbf{p})$ is the proton phase space occupation number
chosen here for simplicity to be the local Fermi distribution,
$f_p(z,b,\mathbf{p})=\Theta(p_{F,p}-|\mathbf{p}|)$;
\begin{equation}
   \sigma_{\bar p p \to R}(s) = \frac{(2J_R+1)\pi}{s/4-m_N^2} 
                                \frac{s \Gamma_{R \to \bar p p}  \Gamma_R}{(s-m_R^2)^2+s\Gamma_R^2}
                                \label{sigma_barpp_to_R}
\end{equation}
is the charmonium formation cross section;    
$v_{\bar p p}=\sqrt{s(s/4-m_N^2)}/E_{\bar p}E_p$ is the $\bar p p$ relative velocity.
We assume the proton energy $E_p=m_N-B$, where $B \simeq 8$ MeV is the nucleus binding energy 
per nucleon. 

The exponential factors in Eq.(\ref{sigma_pbarA2R}) account for the antiproton
and charmonium survival probabilities.
The effective charmonium-nucleon cross section is chosen as
$\sigma_{RN}^{\rm eff}(p_R,z) = \sigma_{RN}(p_R) \kappa(z)$,
where $\sigma_{RN}(p_R)$ is the total interaction cross section of the entirely formed charmonium
with a nucleon; $\kappa(z)$ is the charmonium position dependent function, $\kappa(z) < 1$ for
$z < l_R$ and  $\kappa(z) = 1$ for $z \geq l_R$, which takes into account the charmonium formation
length $l_R$ according to the color diffusion model \cite{Farrar:1989vr}. 
The cross sections $\sigma_{RN}$ have been studied by several groups of authors 
(c.f. \cite{Gerschel:1993uh,Kharzeev:1996yx,Gerland:1998bz,Molina:2012mv}) and are expected to be
within 3.5-7 mb for $J/\psi$ , 0-20 mb for $\psi^\prime$, and 7-16 mb for $\chi_c$-states.

\begin{figure}[ht]
\centering
\includegraphics[scale = 0.4]{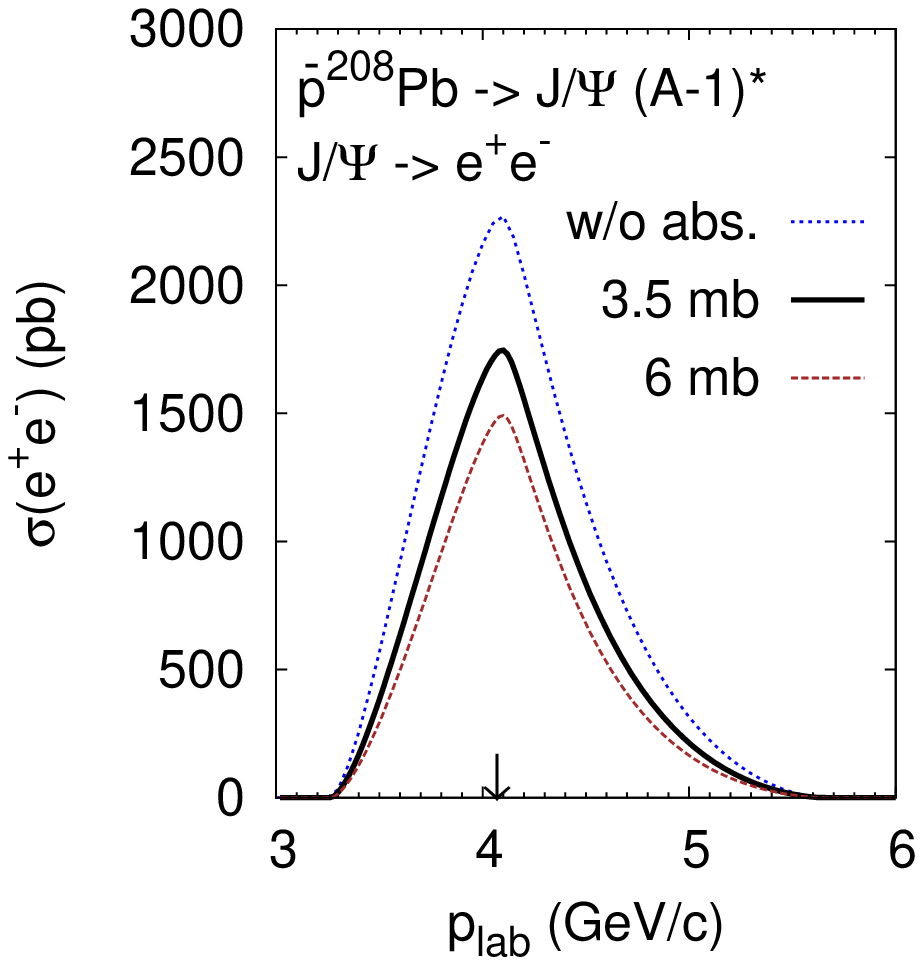}
\includegraphics[scale = 0.4]{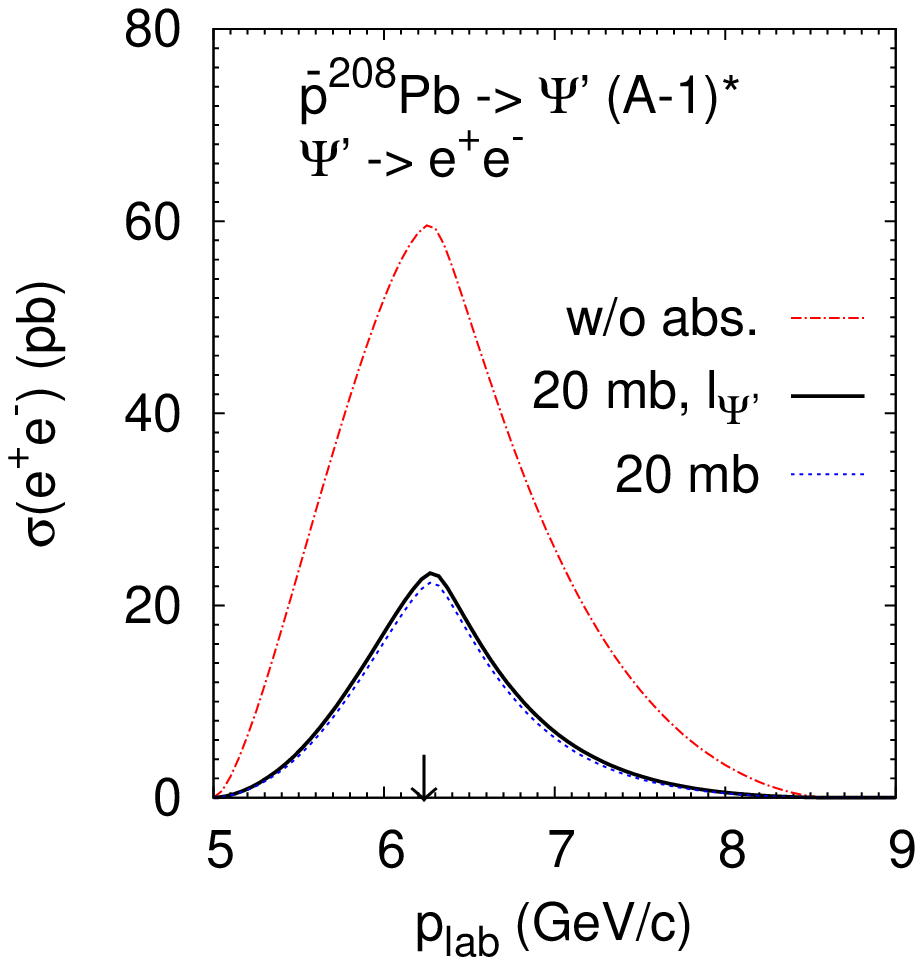}
\caption{$J/\psi$ and $\psi^\prime$ production cross sections on Pb. Arrows show the beam momenta
for the on-shell charmonium production on the proton target: 4.07 GeV/c for $J/\psi$  
and 6.23 GeV/c for $\psi^\prime$.}
\label{fig:sig_vs_plab}
\end{figure}

Fig.~\ref{fig:sig_vs_plab} shows the $J/\psi$ and $\psi^\prime$ production cross sections on 
the lead target as a function of the $\bar p$-beam momentum. We have chosen three representative values of the 
$J/\psi N$ cross section: $\sigma_{J/\psi N}=0,~3.5$ and 6 mb. For the $\psi^\prime N$ cross
section, we considered only the two limiting cases, i.e. $\sigma_{\psi^\prime N}=0$ and 20 mb 
\cite{Gerland:1998bz}. The formation length for $J/\psi$ can be evaluated as
$l_{J/\psi}=2p_{J/\psi}/(m_{\psi^\prime}^2-m_{J/\psi}^2) \simeq 0.1\mbox{fm}~p_{J/\psi}/\mbox{GeV}$.
Thus, at $p_{J/\psi} \simeq p_{\rm lab} \simeq 4$ GeV/c the $J/\psi$ formation length is 0.4 fm which
is even smaller than the internucleon spacing, $d_{NN} \simeq 2$ fm. As a consequence, the $J/\psi$ production
cross sections shown in Fig.~\ref{fig:sig_vs_plab} are almost insensitive to the formation length effects.
For $\psi^\prime$, the formation length is larger ($l_{\psi^\prime} \simeq 2 l_{J/\psi}$, c.f. \cite{Gerland:1998bz}),
although even in this case formation length practically does not influence the production cross sections.  
On the other hand, we observe the sensitivity of the charmonium production cross section to the chosen 
charmonium-nucleon cross section.

\begin{figure}[ht]
\centering
\includegraphics[scale = 0.4]{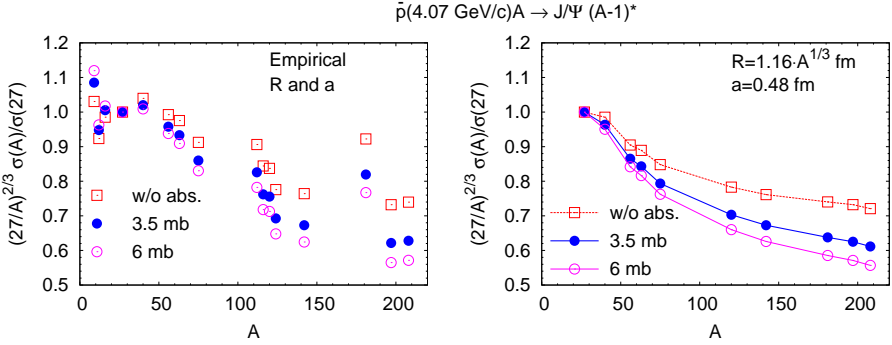}
\caption{Left panel --  transparency ratio, Eq.(\ref{S^tilde_A}), 
for nuclei $^9$Be, $^{12}$C, $^{16}$O, $^{27}$Al,
$^{40}$Ca, $^{56}$Fe, $^{63}$Cu, $^{75}$As, $^{112,116,120,124}$Sn, $^{142}$Ce, 
$^{181}$Ta, $^{197}$Au and $^{208}$Pb calculated with empirical parameters 
of the proton and neutron density profiles.
Right panel -- same for heavy nuclei excluding $^{112,116,124}$Sn 
applying the two-parameter Fermi distribution with radius $R$ and diffuseness $a$ as indicated.}
\label{fig:transRatio}
\end{figure}

Fig.~\ref{fig:transRatio} displays the transparency ratio of the $J/\psi$ production 
\begin{equation}
       \frac{\sigma_{\bar p A \to J/\psi (A-1)^*}}{\sigma_{\bar p\,^{27}{\rm Al} \to J/\psi\,^{26}{\rm Mg}^*}}
       \left(\frac{27}{A}\right)^{2/3}     \label{S^tilde_A}
\end{equation}
at the on-shell peak beam momentum $p_{\rm lab}=4.07$ GeV/c as a function of the target mass number.
The scaling factor $\propto A^{-2/3}$ accounts for the surface absorption of the antiproton.
In general, the transparency ratio is expected to be a more robust observable sensitive to charmonium-nucleon cross section,
since the possible uncertainties in the production width $\Gamma_{\bar p \to R}$ cancel out in Eq.(\ref{S^tilde_A}).
Indeed, we see that the transparency ratio is quite sensitive to the $J/\psi N$ cross section, although for the quantitative
conclusions one should carefully take into account the empirical details of the nuclear density profiles at the surface region.  

\section{$\chi_{cJ} (J=0,1,2)$ production}
\label{chi_c}

Since the mass splitting between the different  $\chi_{cJ}$ states is small ($\sim140$ MeV), 
the nondiagonal transitions $\chi_{cJ_1} N \to \chi_{cJ} N$ should be easily possible. 
The amplitudes of such transitions can be calculated from the Clebsch-Gordan decomposition of a physical  
$\chi_{cJ}(\nu)$ state with helicity $\nu$ in the basis of the  $c \bar c$ states with fixed orbital ($L_z$) and spin ($S_z$)
magnetic quantum numbers \cite{Larionov:2013nga}:
$
    M_{J\nu;J_1\nu}(0)=
     \sum_{L_z,S_z} \langle J\nu |1L_z;1S_z\rangle 
      M_{L_z}(0) \langle 1L_z;1S_z|J_1\nu\rangle   \label{Jnu_M}
$.
With a help of the optical theorem, the forward scattering amplitude $M_{L_z}(0)$ can be expressed via
the total interaction cross section $\sigma_{L_z}$ of a $c \bar c$ pair with fixed $L_z$ with a nucleon:
$
    M_{L_z}(0) = 2 i p_{\rm lab} m_N \sigma_{L_z}
                    (1-i\rho_{\chi N})
$,
with $\rho_{\chi N}=\mbox{Re}M_{L_z}(0)/\mbox{Im}M_{L_z}(0)=0.15-0.30$. In Ref. \cite{Gerland:1998bz}, the cross sections $\sigma_{L_z}$ have been 
calculated by using the nonrelativistic quarkonium model and the QCD factorization theorem. Their values, $\sigma_{\pm1}=15.9$ mb and $\sigma_0=6.8$ mb,
differ by a factor of two, which is the ratio of the transverse size squared for the $c \bar c$ states with $L_z=\pm 1$ and $L_z=0$.
This directly leads to the finite nondiagonal transition amplitudes:
$M_{20;00}(0) = 2 i p_{\rm lab} m_N \frac{\sqrt{2}}{3}(\sigma_1-\sigma_0) (1-i\rho_{\chi N})$,
$M_{2\pm1;1\pm1}(0) = \pm 2 i p_{\rm lab} m_N \frac{1}{2}(\sigma_1-\sigma_0) (1-i\rho_{\chi N})$.

\begin{figure}[ht]
\centering
\includegraphics[scale = 0.4]{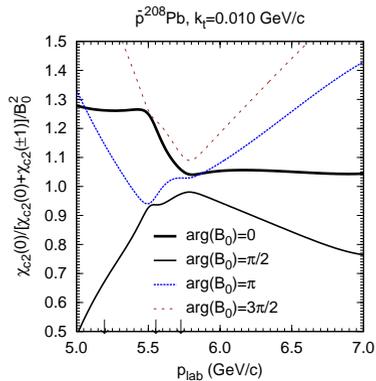}
\caption{Helicity ratio (\ref{calR}) for the different values of the phase difference
of the $\nu=0$ helicity amplitudes for $\chi_{c2}$ and $\chi_{c0}$. Arrows show the beam momenta
for the on-shell charmonium production on the proton target: 5.194, 5.553 and 5.727 GeV/c
for $\chi_{c0}$, $\chi_{c1}$ and $\chi_{c2}$, respectively.}
\label{fig:helRat}
\end{figure}

By applying the generalized eikonal approximation (GEA)(c.f. \cite{Sargsian01} and refs. therein), 
we have calculated the helicity ratio 
\begin{equation}
   {\cal R} = \frac{\chi_{c2}(0)}{[\chi_{c2}(0)+\chi_{c2}(\pm1)]|B_0|^2}   \label{calR}
\end{equation}
for $\chi_{c2}$ production in $\bar p ^{208}$Pb collisions. Here, $B_0$ 
is the helicity amplitude $\bar p p \to \chi_{c2}(\nu=0)$,  $|B_0|^2=0.13\pm0.08$ \cite{Ambrogiani:2001jw}. 
Fig.~\ref{fig:helRat} shows the ratio (\ref{calR}) at small transverse momentum
as a function of $\bar p$-beam momentum. In the absence of any $\chi_{cJ}N$ interactions, 
one has ${\cal R} =1$. However, the interference term of the direct $\bar p p \to \chi_{c2}(0)$
and the two-step $\bar p p \to \chi_{c0}(0)$, $\chi_{c0}(0) N \to \chi_{c2}(0) N$ amplitudes
leads to $\sim 30\%$ deviations of the helicity ratio from one
due to the strong coupling of the $\chi_{c0}$ state with $\bar p p$ channel.    

\section{Summary}
\label{summary}

We have performed the Glauber and the GEA calculations of charmonium production
in $\bar pA$ collisions close to threshold. To summarize the main results:
The strong sensitivity of $J/\psi$ ($\psi^\prime$) production to the
genuine $J/\psi N$ ($\psi^\prime N$) cross sections is confirmed.
For the quantitative determination of the $J/\psi N$ ($\psi^\prime N$) 
cross sections, the density profiles of the target nuclei are important.
Polarization effects in the $\chi_{c2}$ production in the {\it unpolarized}
$\bar p$ collisions with nuclei appear as a consequence of the different transverse
sizes of $c \bar c$ pair with $L_z=\pm1$ and $L_z=0$.
These findings could be useful for the planning of the forthcoming PANDA experiment
at FAIR.  

\bibliography{larionov_MESON2014}

\end{document}